\documentclass[prb,showpacs,showkeys,twocolumn]{revtex4}
\usepackage{graphicx}
\usepackage{amsmath}
\usepackage{bm}

\usepackage{amssymb}

\begin{document}
\title{Surface density of states of $s_\pm$-wave Cooper pairs in a two-band model}
\author{Seiichiro Onari}
\author{Yukio Tanaka}
\affiliation{Department of Applied Physics, and
 JST, TRIP, Nagoya University, 
Chikusa, Nagoya 464-8603, Japan.}

\begin{abstract}
We calculate the surface density of state (SDOS) of $s_\pm$-wave 
Cooper pair in two-band superconductor model, 
where gap functions have different signs 
between two bands. 
We find that the Andreev bound state appears at surface due to 
the sign change in the 
gap function in the interband quasiparticle 
scattering.
However, we do not obtain the zero-energy peak of SDOS 
in contrast to the $d$-wave case. 
The tunneling spectroscopy of $s_\pm$-wave is much more complex as
 compared to the $d$-wave case realized in high-$T_c$ cuprates. 
\end{abstract}

\maketitle

\section{Introduction}
Recent discovery of superconductivity in the iron based
 LaFeAsO$_{1-x}$F$_x$ with $T_c = 26$K\cite{Hosono2} has 
aroused great interests as a 
class of non-cuprate compound.
In the iron-based family, various compounds 
exhibit superconductivity with $T_c$ now exceeding 
55 K. Superconductivity 
has also been found in iron-based materials with
different layered structures that include BaFe$_2$As$_2$\cite{Rotter} and
FeSe\cite{Hsu}.
Local spin-density calculations for LaFeAsO 
have shown that the system 
is around the border between magnetic and nonmagnetic states, with 
a tendency toward antiferromagnetism.\cite{Singh,Xu}  
It has also been pointed out that the 
electron-phonon coupling in this material 
is too weak to account for $T_c=26$K.\cite{Mu,Boeri} 
Based on the first principles calculation, 
minimum five-band model to describe the iron-based superconductor has 
been proposed \cite{Kuroki-Fe}. 
Using this five-band model, pairing symmetry has been calculated 
based on the  random phase approximation (RPA)\cite{Kuroki-Fe}. 
The resulting gap function does not have nodes 
on the Fermi surface while it has a sign change between Fermi surfaces. 
Now, it is called an $s_\pm$-wave pairing \cite{Mazin,Golubov}. 
There have also been relevant theoretical predictions which support
the realization of the $s_\pm$-wave model \cite{Nomura,Ikeda,Yanagi}.
\par
In order to elucidate the energy-gap structure of these
$s_{\pm}$-wave superconductors, 
experiments based on standard technique, $e.g.$, 
NMR\cite{Nakai,Kawabata}, specific heat\cite{Mu}, 
penetration depth\cite{Hashimoto,Malone}
and quasiparticle tunneling spectroscopy have started
\cite{Shan,Yates,Millo,Chen}. 
It is a very challenging issue to clarify 
the superconducting profile of $s_{\pm}$-wave superconductors. 
Since the internal phase degree of freedom exists 
in the gap function of $s_{\pm}$-wave pairing, 
it is natural to expect phase sensitive phenomena realized 
in high-$T_c$ cuprates \cite{dwave1,dwave1-2,dwave1-3,dwave1-4}. 
As shown in the study of high-$T_c$ cuprates, 
the mid gap Andreev bound state (MABS) is formed 
at the surface due to the internal phase effect, where 
a quasiparticle feels a different sign of the gap function
depending on the 
direction of their motions. 
The presence of the MABS produces zero-energy peak (ZEP) 
of the surface density of states (SDOS) 
and has been observed as a zero-bias conductance peak (ZBCP) in tunneling spectroscopy 
up to now \cite{dwave,dwave-2,dwave-3,Tanuma}. 
It is an urgent topic to reveal whether 
MABS exists in the $s_{\pm}$-wave pairing or not. \par
Several theories of surface or 
interface profiles about $s_{\pm}$-wave pairing
have been presented very recently 
 \cite{Bang,Ghaemi,Choi,Linder,Tsai,Feng,Golubov2,Nagai}. 
However, SDOS of two-band superconductors 
with $s_{\pm}$ model has not been understood yet. 
The presence or absence of MABS have not been resolved yet. 
Furthermore, character of the nonzero inner gap Andreev bound state
(ABS) has not been clarified. 
To reply to these issues, in the present paper, 
we employ a simple two-band tight-binding model with 
$s_\pm$-wave as a prototype of iron-based pnictides, and calculate 
SDOS  for the [100] and [110] oriented interfaces using the
t-matrix  method\cite{Matsumoto}. 
A merit of our  calculation is that 
details of the band structure and band mixing can be 
 microscopically taken into account.
We find that ABS 
with nonzero energy is formed at the 
surface due to the interband quasiparticle scattering,   
through which gap functions change sign. 
However, there is no ZEP in SDOS in contrast to the case of $d$-wave pairing realized in high-$T_c$ cuprates\cite{dwave,dwave-2,dwave-3}.

\section{Model and formulation}
\begin{figure}
\begin{center}
\includegraphics[width=8cm]{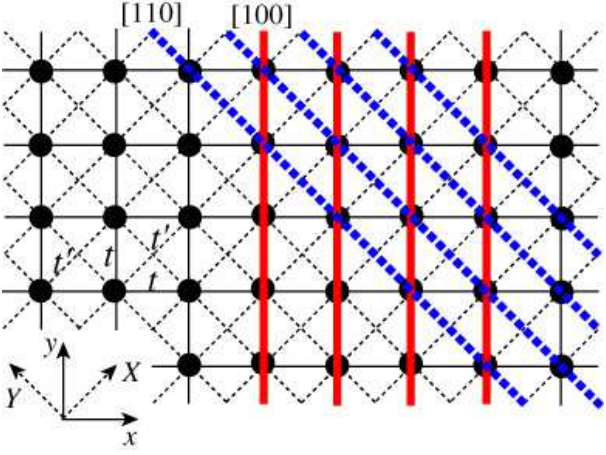}
\caption{(Color online) Two-dimensional square lattice with nearest-neighbor hopping
 $t$ (thin solid line) and next-nearest-neighbor hopping $t'$ (thin 
 dotted line). The [100] and [110] oriented surfaces are constructed by 
 inserting four infinite potential barriers illustrated with 
thick solid lines and  thick dotted lines, respectively}
\label{lattice}
\end{center}
\end{figure}

We start with a two-band tight-binding model on a square lattice. 
Since there has not been an explicit study about the SDOS of the $s_{\pm}$ model, we
choose the energy dispersion of two orbitals simply 
supposing $d_{xz}$ and $d_{yz}$ orbitals or $p_{x}$ and $p_{y}$
orbitals.
Hereafter, we define index $1$ in matrix form as $d_{xz}$ ($p_x$)
orbital and index 2 as $d_{yz}$ ($p_y$) orbital.
$X$ and $Y$ axes are rotated by 
45 degrees from $x$-$y$, where $x$ and $y$ denote the axes in a unit cell as shown in Fig. \ref{lattice}.

\begin{figure}
\begin{center}
\includegraphics[width=8cm]{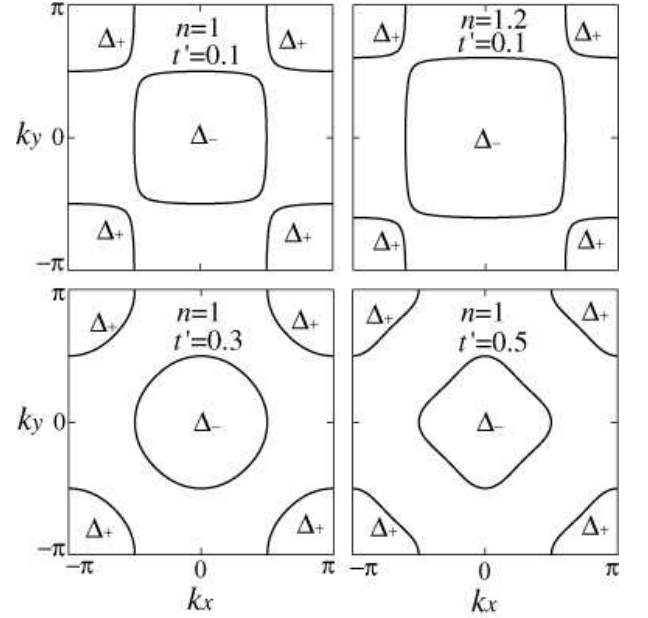}
\caption{Outer (inner) Fermi surfaces with gap function $\Delta_{+(-)}$
 consist of band $+(-)$.}
\label{fermi}
\end{center}
\end{figure}

First, we discuss the normal state. 
Tight-binding Hamiltonian is given in the form 
\begin{eqnarray}
H_0=\sum_{ij}\sum_{\mu\nu}\sum_\sigma
t_{i\mu,j\nu}c_{i\mu\sigma}^\dagger c_{j\nu\sigma},
\end{eqnarray}
where $t_{i\mu,j\nu}$ is a hopping integral from the $\nu$-th orbital on
the $j$-th 
site to the $\mu$-th orbital on the $i$-th site, 
$c_{i\mu\sigma}^\dagger$ creates an electron with spin $\sigma$ 
on the $\mu$-th orbital at site $i$.
As shown in Fig. \ref{lattice},  we take the nearest-neighbor hopping $t$
and the next-nearest-neighbor hopping $t'$.
The band filling $n$ is defined as the number of electrons per number of sites
(e.g., $n=1$ for half filling). 
The Hamiltonian in the Fourier transformed representation is given as
\begin{equation}
H_0=\sum_{\bm{k}}\sum_{\mu\nu}\sum_\sigma
\hat{\varepsilon}^0_{\mu\nu}(\bm{k})c_{\bm{k}\mu\sigma}^\dagger
c_{\bm{k}\nu\sigma},
\end{equation}
where the $2\times 2$ matrix $\hat{\varepsilon}^0(\bm{k})$ is denoted by
\begin{equation}
\hat{\varepsilon}^0(\bm{k})=\left(
\begin{array}{cc}
-t\cos k_x & 2t'\sin k_x\sin k_y\\
2t'\sin k_x\sin k_y & -t\cos k_y
\end{array}
\right).
\end{equation}
Hereafter, we take $t$ and the lattice constant $a$ as the units for
energy and length, respectively.
$\hat{\varepsilon}^0(\bm{k})$ can be diagonalized to $\varepsilon^0_a(\bm{k})$,
which corresponds to the energy of band $a$
\begin{equation}
\varepsilon^0_a(\bm{k})=\sum_{\mu\nu}U^*_{\mu a}(\bm{k})U_{\nu a}(\bm{k})\hat{\varepsilon}^0_{\mu\nu}(\bm{k}),
\end{equation}
where $U(\bm{k})$ is a $2\times 2$ unitary matrix.
Fermi surfaces consist of two parts near the half filling as shown
in Fig. \ref{fermi}.
We define the band which forms inner (outer) Fermi surface as band $-(+)$.

In a two-band model, gap function generally forms a $2\times 2$
matrix. The gap function $\hat{\Delta}_{\mu\nu}(\bm{k})$ in the orbital
representation  is
transformed to the gap function $\Delta_{ab}(\bm{k})$ in the band representation using the unitary matrix $U(\bm{k})$, 
\begin{equation}
\Delta_{ab}(\bm{k})=\sum_{\mu\nu}U^*_{\mu a}(\bm{k})U^*_{\nu b}(\bm{-k})\hat{\Delta}_{\mu\nu}(\bm{k}).
\end{equation}
Here, we neglect the frequency $\omega$ dependence of the gap function and assume
that the gap function in the band representation is
diagonal, $\Delta_a(\bm{k})=\Delta_{aa}(\bm{k})$. 
Gap function of band $-$ (inner Fermi surface) and band $+$ (outer Fermi
surface) are denoted by $\Delta_-$ and
$\Delta_+$, respectively as shown in Fig. \ref{fermi}.
Then, the gap function in the orbital representation is obtained as
\begin{equation}
\hat{\Delta}_{\mu\nu}(\bm{k})=\sum_{a}U_{\mu a}(\bm{k})U_{\nu a}(\bm{-k}){\Delta}_{a}(\bm{k}).
\end{equation}
In the case that the hopping integral is given by a real number, the relation
$\hat{\varepsilon}^0_{\mu\nu}(\bm{k})=\hat{\varepsilon}^{0*}_{\mu\nu}(-\bm{k})$
is satisfied. 
Then, we take following relation
\begin{equation}
U_{\mu a}(\bm{k})=U^*_{\mu a}(-\bm{k}).
\end{equation}
Using the above gap function, 
bulk Green's function $\hat{G}(\omega,\bm{k})$ in the
superconducting state is given by a
$4\times4$ Nambu representation as follows 
\begin{equation}
\hat{G}(\omega,\bm{k})=\left[\omega-\left(
\begin{array}{cc}
\hat{\varepsilon}^0(\bm{k})-\mu & \hat{\Delta}(\bm{k})\\
\hat{\Delta}^\dagger(\bm{k}) & -\hat{\varepsilon}^0(\bm{k})+\mu
\end{array}
\right)\right]^{-1},
\end{equation}
with chemical potential $\mu$. 
In the actual numerical calculation, we replace 
$\omega$ by $\omega + i \gamma$ with small real number 
$\gamma$ to avoid divergence of the integral. 
Local density of states (LDOS)
in the bulk 
is obtained by 
$-1/(N\pi)\sum_{\bm{k},l=1,2}{\rm Im}\hat{G}_{ll}(\omega,\bm{k})$, where
$N$ denotes $\bm{k}$-point meshes.
The Green's function of the inhomogeneous system 
including surface $\hat{G}^{s}(\omega,\bm{r},\bm{r}')$ is 
calculated by 
$\hat{G}(\omega,\bm{r})$, which is the
Fourier transformed form of $\hat{G}(\omega,\bm{k})$. 
As shown in Fig. \ref{lattice}, we insert the infinite 
potential barrier $Z(\bm{r})$ in four-atomic layers parallel 
to the surface in the actual calculation. 
$\hat{G}^{s}(\omega,\bm{r},\bm{r}')$ is given by 
\begin{eqnarray}
\hat{G}^s(\omega,\bm{r},\bm{r}')&=&\hat{G}(\omega,\bm{r}-\bm{r}')\nonumber\\
&+&\int d\bm{r}''\hat{G}(\omega,\bm{r}-\bm{r}'')Z(\bm{r}'')\hat{\tau}_3\hat{G}^s(\omega,\bm{r}'',\bm{r}'),
\end{eqnarray}
where $\hat{\tau}$ denotes the Pauli matrix in charge space. We note that
$\hat{G}^s$ breaks translational symmetry.
Using the $\hat{G}^s(\omega,\bm{r},\bm{r}')$, we obtain SDOS by
$-1/\pi\sum_{l=1,2}{\rm Im}\hat{G}^s_{ll}(\omega,\bm{r}_s,\bm{r}_s)$,
where $\bm{r}_s$ denotes the location of the surface.
Throughout this study, we take $N=4096\times4096$ $\bm{k}$-point meshes
and $\gamma=0.003$. 

\section{Result}
\begin{figure}
\begin{center}
\includegraphics[width=8cm]{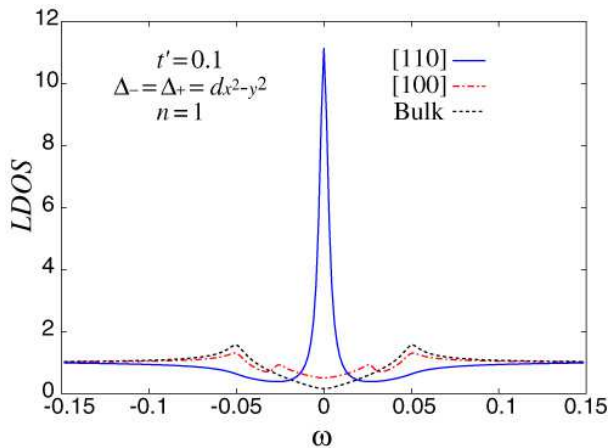}
\caption{(Color online) SDOS of the [110] and [100] oriented surfaces and LDOS in  bulk
 are depicted with solid line, dashed-dotted line and dotted line,
 respectively for $n=1$, $t'=0.1$, and $\Delta_-=\Delta_+=0.05(\cos
 k_x-\cos k_y)$ ($d_{x^2-y^2}$-wave). These values are normalized by the 
values in the normal state. }
\label{dosd}
\end{center}
\end{figure}

In the following, we focus on LDOS
at the surface, i.e., SDOS and bulk. 
First we show the result of the $d_{x^2-y^2}$-wave case for $n=1$, $t'=0.1$ in
Fig. \ref{dosd}, where the gap function of band $+(-)$ is chosen as 
$\Delta_{+(-)}=0.05(\cos k_x-\cos k_y)$. 
Throughout the present study, 
LDOS is normalized to that of the value in the normal state 
at $\omega=0$. 
We see a sharp ZEP of SDOS in 
[110] oriented surface and V-shaped LDOS in the bulk, which are 
consistent with the case of the single band 
$d_{x^2-y^2}$-wave\cite{dwave,dwave-2,dwave-3}. 
The origin of the sharp ZEP is 
the sign change in the gap function felt by quasiparticles 
scattered at the surface, where the momentum of the 
quasiparticles parallel to the [110] surface is conserved 
\cite{dwave,dwave-2,dwave-3}. 

\begin{figure}
\begin{center}
\includegraphics[width=8cm]{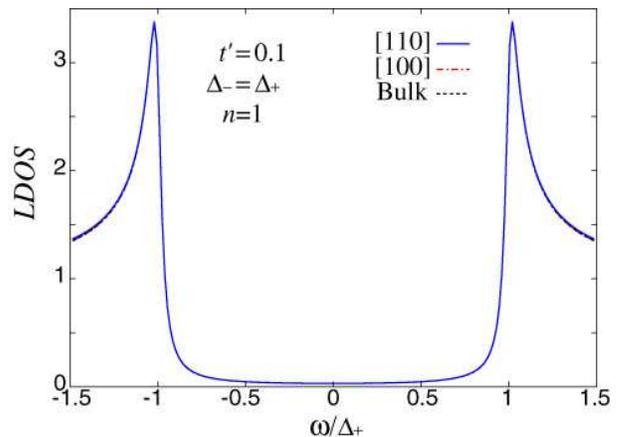}
\caption{(Color online) Plots of LDOS similar to Fig. \ref{dosd} for $n=1$, $t'=0.1$,
 and $\Delta_-=\Delta_+=0.1$ ($s$-wave).}
\label{doss}
\end{center}
\end{figure}

As a reference, we show the result of $s$-wave pairing 
in Fig. \ref{doss}, where
$n=1$, $t'=0.1$ and $\Delta_-=\Delta_+=0.1$. 
The line shapes of SDOS of the [100] and [110] oriented surfaces, 
and LDOS of bulk are almost identical. 
This behavior is robust irrespectively of the band structure 
as far as the relation $\Delta_-=\Delta_+$ is satisfied.

\begin{figure}
\begin{center}
\includegraphics[width=8cm]{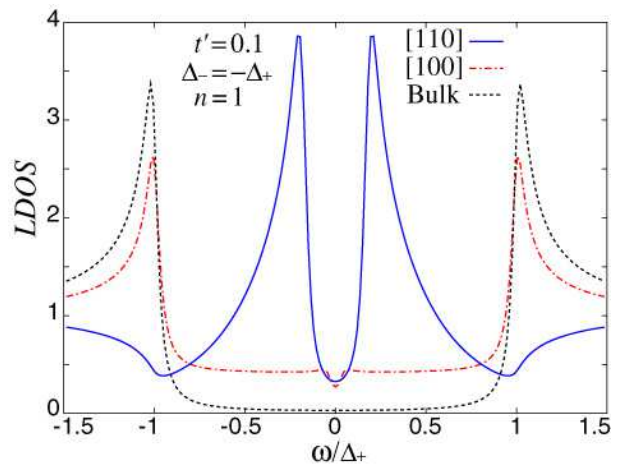}
\caption{(Color online) Plots of LDOS similar to Fig. \ref{dosd} for $n=1$, $t'=0.1$,
 and $\Delta_-=-\Delta_+=-0.1$ ($s_\pm$-wave).}
\label{dos1}
\end{center}
\end{figure}

Next, we move to the $s_\pm$-wave case.
The corresponding results 
for $s_\pm$-wave with $n=1$, $t'=0.1$ are shown in 
Fig. \ref{dos1}, where we choose $\Delta_{+(-)}=+(-)0.1$.
We see that two sharp peaks within the bulk energy gap 
appear in the [110] oriented 
SDOS \cite{Ghaemi}. 
The clear difference from Fig.3 is that there is no ZEP.
For the [100] oriented surface, the value of the 
corresponding SDOS within the gap is almost constant with 
nonzero value. 

\begin{figure}
\begin{center}
\includegraphics[width=8cm]{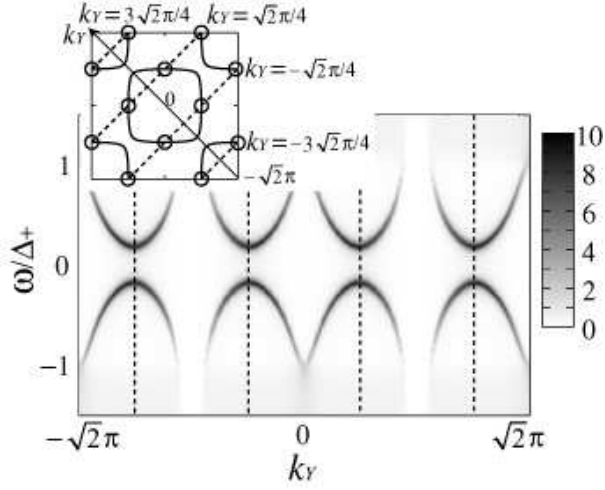}
\caption{Contour plot of the $k_Y$-resolved LDOS(SDOS) in the [110] oriented surface
 for $n=1$, $t'=0.1$, and $\Delta_-=-\Delta_+=-0.1$ ($s_\pm$-wave),
 where $k_Y=\pm\sqrt{2}/4$ and $k_Y=\pm 3\sqrt{2}/4$ are depicted as
 dotted lines. In the inset circles denote points on the Fermi surface with
 $k_Y=\pm\sqrt{2}/4$ and $k_Y=\pm 3\sqrt{2}/4$, which mainly contribute
 to the LDOS.}
\label{dosk}
\end{center}
\end{figure}

In order to clarify the origin of the two peaks in SDOS of the [110]
oriented surface, we show the $k_Y$-resolved 
SDOS of the [110] surface in Fig. \ref{dosk}, 
where $k_Y$ denotes the momentum
parallel to the [110] surface. 
The $k_Y$-resolved SDOS is enhanced at 
$k_Y\sim \pm\sqrt{2}\pi/4,\pm3\sqrt{2}\pi/4$, which
correspond to
$(k_x,k_y)=(0,\pm\pi/2),(\pm\pi/2,0),(\pm\pi,\pm\pi/2),(\pm\pi/2,\pm\pi)$
at the original Fermi surface as shown in the inset of Fig. \ref{dosk}.
At these points, angle resolved 
SDOS has a large value.  
Furthermore, quasiparticles feel a different sign of the gap function
through the scattering between inner and outer bands, which brings about
the ABS. The large momentum change in quasiparticles is automatically
induced by infinite potential barriers inserted at the surface due to
normal (backward) reflection.

On the other hand, the $k_y$-resolved SDOS in the [100] surface (not shown) is
enhanced at $k_y=\pm\pi/2$ within the gap.
Since scattering of the quasiparticle  
preserving $k_y=\pm\pi/2$ occurs between the inner and
 outer Fermi surfaces, 
ABS appears within the bulk energy gap, which makes
residual LDOS, shown as the dashed-dotted line in Fig. \ref{dos1}.

\begin{figure}
\begin{center}
\includegraphics[width=8cm]{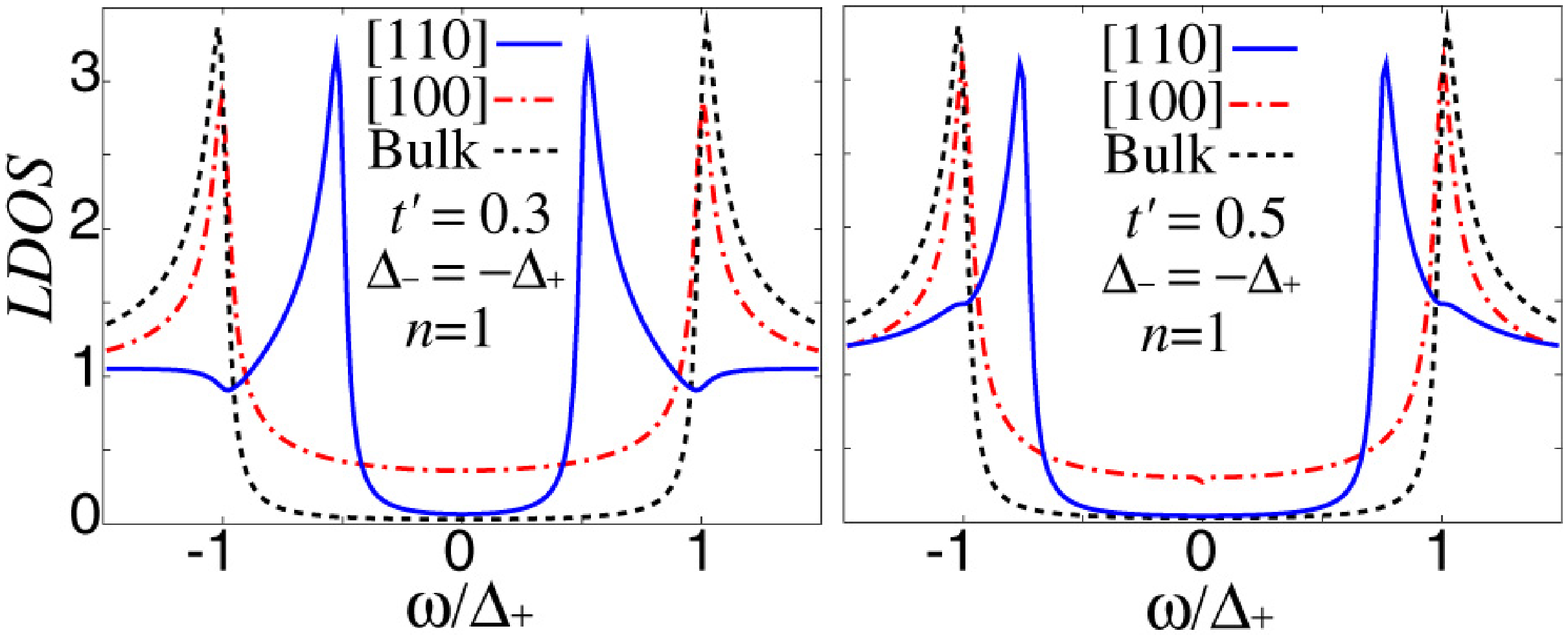}
\caption{(Color online) Plots of LDOS similar to Fig. \ref{dosd} for $n=1$, $\Delta_-=-\Delta_+=-0.1$, $t'=0.3$ (left panel) and
 $t'=0.5$ (right panel).}
\label{dos2}
\end{center}
\end{figure}

Although the sign change in the gap function 
does not produce the ZEP as in the 
case of unconventional superconductors 
such as $d$- or $p$-wave pairing, 
the sign change in the gap function felt by quasiparticle 
enhances the magnitude of the inner gap LDOS. 
In a certain case, it produces sharp peaks, 
shown as a solid line in Fig. \ref{dos1}. 
In order to confirm whether the above behaviors are robust or not, we
change the shape of the Fermi surface 
by controlling the value of
$t'$ for $n=1$, $\Delta_-=-\Delta_+=-0.1$. 
As shown in Fig. \ref{dos2}, the 
positions of the two peaks in SDOS for the [110] oriented 
surface (solid line) 
move toward that of bulk LDOS (dotted line)
 as the value of $t'$ increases. 
Thus, we find that the positions of peaks of the [110] oriented 
SDOS are sensitive to
the shape of the Fermi surface. 
The resulting peak positions are relevant to the 
relative position between the outer and inner Fermi surface.


\begin{figure}
\begin{center}
\includegraphics[width=8cm]{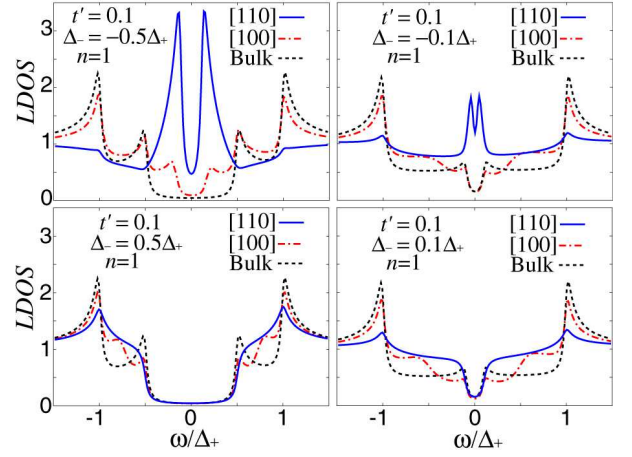}
\caption{(Color online) Plots of LDOS similar to Fig. \ref{dosd} for $n=1$, $t'=0.1$,
 $\Delta_-=-0.5\Delta_+$ (top left panel),
 $\Delta_-=-0.1\Delta_+$ (top right panel), $\Delta_-=0.5\Delta_+$
 (bottom left panel), and 
 $\Delta_-=0.1\Delta_+$ (bottom right panel).}
\label{dos3}
\end{center}
\end{figure}

It is also interesting to clarify how the above results are 
influenced by changing the value of $\Delta_{-}$. 
In Fig. \ref{dos3}, we focus on the $\Delta_-$ dependence 
of SDOS and bulk LDOS for $\Delta_+=0.1$, $n=1$ and
$t'=0.1$. 
Two-gap structure appears in bulk LDOS (dotted lines).  
It is very clear that the resulting bulk LDOS is 
insensitive to the sign of $\Delta_{-}$ 
by comparing the $\Delta_{-}=0.5\Delta_{+}$
($\Delta_{-}=0.1\Delta_{+}$) case  
with the $\Delta_{-}=-0.5\Delta_{+}$ 
($\Delta_{-}=-0.1\Delta_{+}$) case. 
As far as we are looking at bulk LDOS, 
there is no difference between $s_{\pm}$-wave and 
$s$-wave. 
The internal phase degree of the gap function does not appear 
in the bulk LDOS.  
On the other hand, sharp peaks of ABS appear only for 
SDOS of the [110] oriented surface with
$\Delta_{-}=-0.5\Delta_{+}$ and $\Delta_{-}=-0.1\Delta_{+}$. 
The position of the sharp peaks moves toward $\omega=0$ with the 
decrease of the magnitude of $\Delta_{-}$.  
At the same time the height of the peaks is reduced
(solid lines in the upper two panels). 
On the other hand, SDOS for the [100] oriented surface 
does not have clear peaks as compared to that for 
the [110] oriented surface. 
%

\begin{figure}
\begin{center}
\includegraphics[width=8cm]{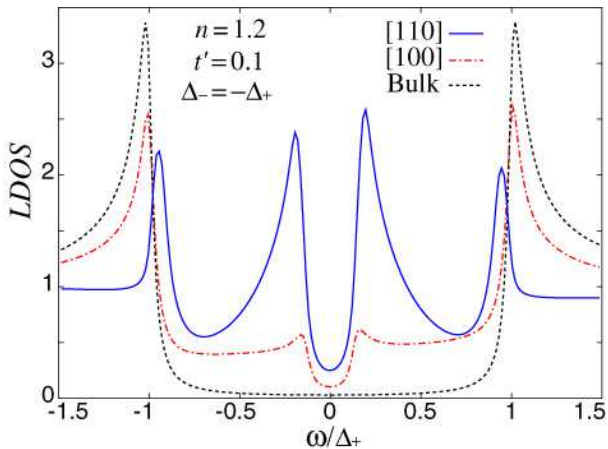}
\caption{(Color online) Plots of LDOS similar to Fig. \ref{dosd} for
 $n=1.2$, $t'=0.1$, $\Delta_-=-\Delta_+=-0.1$.}
\label{dos4}
\end{center}
\end{figure}

\begin{figure}
\begin{center}
\includegraphics[width=8cm]{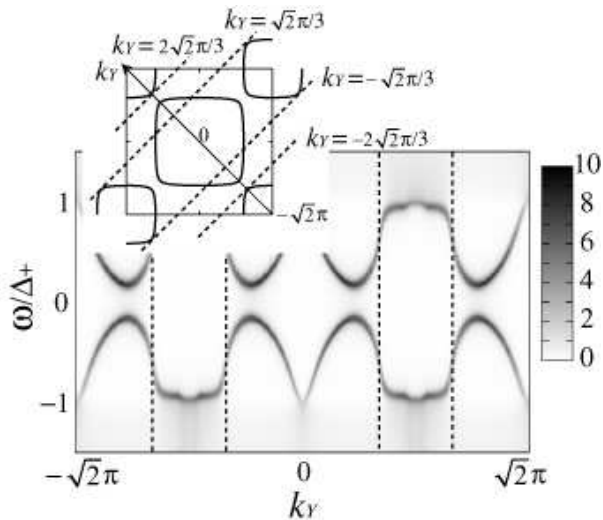}
\caption{Contour plot of the $k_Y$-resolved LDOS(SDOS) in the [110] oriented surface
 for $n=1.2$, $t'=0.1$, and $\Delta_-=-\Delta_+=-0.1$, where $k_Y=\pm\sqrt{2}/3$ and $k_Y=\pm 2\sqrt{2}/3$ are depicted as
 dotted lines. In the inset we see that there is no interband
 pair scattering for $-2\sqrt{2}\pi/3<k_Y<-\sqrt{2}\pi/3$ and 
$\sqrt{2}\pi/3<k_Y<2\sqrt{2}\pi/3$.}
\label{dosk2}
\end{center}
\end{figure}

Finally, we show the result of $n=1.2$ for $t'=0.1$ and 
$\Delta_-=-\Delta_+=-0.1$ in Fig. \ref{dos4}.
In this case, SDOS for the [110] oriented surface 
has two peaks at $\omega=\pm\Delta_+$ 
in addition to the two inner gap peaks. 
This is understood by using 
the $k_Y$-resolved SDOS in Fig. \ref{dosk2}.
We see 
that ABS within gap 
vanishes for $-2\sqrt{2}\pi/3<k_Y<-\sqrt{2}\pi/3$ and 
$\sqrt{2}\pi/3<k_Y<2\sqrt{2}\pi/3$ 
 since interband pair scattering is prohibited due to the absence of
 outer Fermi surface as shown in the inset of Fig. \ref{dosk2}. 
Thus, in this $k_{Y}$ region, the angle-resolved LDOS is 
an independent summation of LDOS in the
outer and inner Fermi surface.  
Then, the resulting angular averaged LDOS has 
peaks at $\omega=\pm \Delta_{+}$.

Summarizing the above results, although the quasiparticle feels 
a sign change of the gap function for fixed $k_{Y}$ ($k_{y}$) 
in the reflection process at the surface, 
ZEP does not appear in the SDOS of $s_\pm$-wave pairing. 
Thus, ZBCP in tunneling spectroscopy 
appears neither the [100] nor the [110] oriented
junctions. 
These features are completely 
different from the $d$-wave gap function in high-$T_c$ cuprates 
where ZBCP appears for the [110] orientation\cite{dwave,dwave-2,dwave-3,Matsumoto}.
One of the big difference from the $d$-wave case 
is that quasiparticles do not feel 
the sign change in the 
gap function as far as it is scattered within 
the same band in $s_{\pm}$-wave pairing. 
There are always both intraband and interband pair scattering. 
The presence of intraband scattering without a sign change 
may prohibit the generation of MABS and ZEP of SDOS. 
In the present study, we employ a simple two-band model.
We admit that five orbitals are needed to describe 
the superconductivity of iron-based superconductors.  
Our final goal is to establish a theory of tunneling spectroscopy 
/ surface density of state taking into account five bands. 
However, up to now, there has not been fully microscopic theory of 
surface density of states of multi-band superconducting systems even in
two-bands cases. To understand the essence of the interference effects
originating from the existence of the multiband, it is reasonable to
start with the two-band model. 
Thus, in the present paper, we have chosen the two-band model for the 
first step as a prototype of multiband model. 
In the near future, we will report the results 
based on the more realistic five-band model. \par
Finally, we comment about the relevance of the present 
$s_{\pm}$-wave model in two-band systems and 
two-band model in non-centrosymmetric 
superconductors\cite{Iniotakis}. 
Recently, there are several studies about surface density of states of non-centrosummetric superconductors. 
The presence of the Rashba spin-orbit coupling induces Fermi surface
splitting, and a similar situation in the present two-band model seems to be realized. 
Due to the presence of Rashba interaction, 
the spatial inversion symmetry is broken in these systems. 
Then spin-singlet $s$-wave and spin-triplet $p$-wave 
pairing can mix each other. 
If the magnitude of the $p$-wave component is larger than that of $s$-wave one, ABS exists and MABS is possible 
for the perpendicular injection of the quasiparticle. 
The resulting ABS can be regarded as  helical edge modes and 
carry spin current. The direction of the current 
flow corresponding to each Kramers doublet is opposite. 
On the other hand, the profile of the ABS 
in the present 
two-band $s_{\pm}$-wave model is very different. 
In the present case, there is no spin current and spin degeneracy remains.

\section{Conclusions}
In this paper, we have calculated the [100] and the [110] 
oriented SDOSs 
for $s_{\pm}$-pairing in a two-band model by changing the 
shape of the Fermi surface and 
the band filling. 
It has been revealed that the inner gap sharp peaks appear 
for SDOS in the [110] oriented surface.  
These peaks originate from the 
ABS
caused by the interband scattering of quasiparticles, 
through which gap functions change sign. 
Such  sharp peaks do not appear in the $s$-wave case, where 
there is no sign change in the gap function between the two bands. 
It is also noted that the  resulting SDOS of $s_{\pm}$ model 
does not have ZEP.
This means that the tunneling spectroscopy of $s_\pm$ 
superconducting state is much more complex as compared to 
the $d$-wave case realized in high-$T_c$ cuprates. \par
Up to now, there has been experimental reports about tunneling spectroscopy. 
The experimental line shapes of tunneling conductance 
are distributed including  
gap structures\cite{Kawabata,Hashimoto,Malone,Chen}
 and ZEP\cite{Shan,Yates,Millo}. 
However, the experimental condition has not been clarified yet up to now. 
In the light of the study of high $T_{c}$ cuprate \cite{Alff,Iguchi}, 
it has been revealed that well-oriented surface or well-oriented
interface with low transparency junctions are needed
to compare the surface density of states with the actual tunneling
conductance \cite{dwave-tunnel}. 
We hope tunneling spectroscopy of well-oriented surface 
or well-controlled junctions with low transparency will be 
attainable in the present iron-based superconductors by the  
progress of microfabrication technique. \par
There are several interesting future problems. 
In the present paper, we have solved the Green's function in tight-binding model.
It is possible to solve the Bogoliubov de-Gennes equation in the lattice model. 
The study along this direction is useful to elucidate interference effect 
much more in detail \cite{Tanuma1,Tanuma2}. 
Josephson effect in the $s_{\pm}$-wave superconductor may be 
fascinating since we can detect internal phase effect 
\cite{Ohashi,Machida}. It is interesting to clarify the possible 
existence of nonmonotonic temperature dependence in 
high-$T_c$ cuprate junctions \cite{J1,J2}. 
Proximity effect in $s_{\pm}$-wave superconductors is also an interesting 
topic. Through the study of the proximity effect in unconventional 
superconductors \cite{Proximity}, 
the odd-frequency pairing amplitude has a crucial role to 
characterize the bound state \cite{oddfrequency}. 
It is a challenging issue to clarify the induced odd-frequency pairing 
near the present two-band model.

%
\par

{\bf Acknowledgments}\\
We are grateful to A. A. Golubov and Y. Nagai for useful comments and discussions.
Numerical calculations have been performed at the facilities of
the Information Technology Center, University of Tokyo, 
and also at the Supercomputer Center,
ISSP, University of Tokyo. 
This study has been supported by 
Grants-in-Aid for the 21st Century COE
``Frontiers of Computational Science.''

\end{document}